\documentclass[twocolumn,showpacs,preprintnumbers,aps]{revtex4}
\usepackage{epsf}
\usepackage{dcolumn}
\usepackage{bm}
\draft

\begin{document}
\title{Testing of T-odd, P-even interactions by nonpolarized neutron transmission through
a nonpolarized nuclear target placed into electric field}
\author{ V. G. Baryshevsky}
\affiliation{Research Institute for Nuclear Problems, Belarusian State University,
11 Bobryiskaya str., 220050, Minsk, Republic of Belarus,\\
E-mail: bar@inp.minsk.by }
\date{\today}

\begin{abstract}
A new possibility for the study of time-reversal violation is described. 
It consists in measurement of nonpolarized neutron transmission through nonpolarized 
nuclear target placed into electric field
\end{abstract}


\maketitle

\narrowtext

During the last years many theoretical and experimental works were
devoted to the study of symmetry violation in hadronic systems.

The five-fold correlation (FC) test searches for T-odd P-even term in the
forward elastic scattering amplitude of the form 
$(\overrightarrow{s}[\overrightarrow{k} \times \overrightarrow{J}])(\overrightarrow{k} \overrightarrow{J})$,
$\overrightarrow{s}$ and $\overrightarrow{k}$ are the neutron spin and momentum, $\overrightarrow{J}$
is the target spin [1-5].

According to \cite{3} the experimental limits for T-odd P-even interactions in
hadronic systems are only $10^{-3}$ with respect to the conventional T-even P-even
strong interaction.

One of the difficulties in increasing the sensibility of experiments studying FC phenomena
is the necessity of use of an aligned nuclear target (making of an aligned target requires cooling 
that restricts available target size). 

Additional challenge is a need of a beam of polarized neutrons (protons and so on).

This letter considers appearance of a T-odd P-even term in amplitude of forward scattering of a
neutron (proton) by a nucleus in case of the target placing in an external electric field
$\overrightarrow{E}$:
\begin{equation}
f_{E}(0)=D \overrightarrow{k} \overrightarrow{E},
\label{amplitude}
\end{equation}

As a result the amplitude of forward scattering of a particle with spin $\frac{1}{2}$ by a 
nonpolarized nucleus can be expressed as:
\begin{equation}
f(0)=A+B{\overrightarrow{s}\overrightarrow{k}}+C{\overrightarrow{s}\overrightarrow{E}}
+D{\overrightarrow{k}\overrightarrow{E}}+F{\overrightarrow{s}
[\overrightarrow{k} \times \overrightarrow{E}]}
\label{f0}
\end{equation}
terms $A$ and $F$ describes conventional T-even P-even interactions, $B$ and $C$ describe the well-known
P-odd T-even and T-odd P-odd contributions, respectively, while $D$ corresponds T-odd P-even interaction.
Suppose $\overrightarrow{k}$ and $\overrightarrow{E}$ are parallel then term with $F$ is absent.
It is known that the term proportional to $\overrightarrow{s}\overrightarrow{k}$ is responsible for
P-odd rotation of the neutron spin around $\overrightarrow{k}$ direction and spin dichroism i.e. difference
of neutron absorption cross-section for neutrons with the spin parallel and antiparallel to $\overrightarrow{k}$.
And similarly, the term proportional to $\overrightarrow{s}\overrightarrow{E}$ describes T-odd P-odd
rotation of the neutron spin around $\overrightarrow{E}$ direction and spin dichroism i.e. difference
of neutron absorption cross-section for neutrons with the spin parallel and antiparallel to $\overrightarrow{E}$.
This similarity provides to apply for T-odd P-odd experiment $\sim \overrightarrow{s}\overrightarrow{E}$ 
the well developed methods for study the T-odd P-even effects 
caused by $\overrightarrow{s}\overrightarrow{k}$.

It is important that the T-odd P-odd contribution to $f(0)$ 
described by the term $\overrightarrow{s}\overrightarrow{E}$
does not require target polarization in contrast with the
{widely discussed in literature}
term proportional to $\overrightarrow{s}[\overrightarrow{k} \times \overrightarrow{J}]$ \cite{6,7}.

The term proportional to $\overrightarrow{k} \overrightarrow{E}$ describes the
T-odd P-even contribution to the refractive index. The presence of this term
makes the phase of a neutron wave $\varphi$ after passing through a layer of width $L$ as:
\begin{equation}
\varphi=k~Re(n-1)L
\end{equation}
depending on the mutual orientation of $\overrightarrow{k}$ and $\overrightarrow{E}$
\begin{equation}
\Delta \varphi=\varphi(\overrightarrow{k} \uparrow \uparrow \overrightarrow{E})-
\varphi(\overrightarrow{k} \uparrow \downarrow \overrightarrow{E})=4 \pi \rho ~Re D E.
\end{equation}

The above phase difference
 could be measured either by a neutron interferometer or 
considering neutron diffraction in a crystal.

The imaginary part of the refractive index, being proportional to $Im~D$,
describes T-odd P-even dependence of total scattering cross-section 
$\sigma_{tot}$ on
the mutual orientation of $\overrightarrow{k}$ and $\overrightarrow{E}$
(according to the optical theorem $Im f(0)=\frac{k}{4 \pi} \sigma_{tot}$).

From (\ref{f0}) it follows that considering passing of nonpolarized neutrons
through a nonpolarized target one can express the total cross-section of
scattering of a nonpolarized neutron by a nonpolarized nucleus as:
\begin{equation}
\sigma_{tot}=\sigma_{0}+\sigma_{E}\overrightarrow{n}_{k}\overrightarrow{n}_{E}
\end{equation}
where $\overrightarrow{n}_{k}$ and $\overrightarrow{n}_{E}$ are the unit vectors,
$\overrightarrow{n}_{k}=\frac{\overrightarrow{k}}{k}$ and 
$\overrightarrow{n}_{E}=\frac{\overrightarrow{E}}{E}$.
Therefore,   
the intensity of a neutron beam passed trough a target 
of $L$ width can be written as:
\begin{eqnarray}
{\text for } \overrightarrow{n}_{k}\uparrow \uparrow \overrightarrow{n}_{E},~ 
J_{\uparrow \uparrow}=J_0 e^{-\rho (\sigma_{0}+\sigma_{E})L} \\
{\text for } \overrightarrow{n}_{k}\uparrow \downarrow \overrightarrow{n}_{E},~ 
J_{\uparrow \downarrow}=J_0 e^{-\rho (\sigma_{0}-\sigma_{E})L},
\end{eqnarray}
where $J_0$ is the intensity of the incident neutron beam and $\rho$ is the
number of nuclei in cm$^3$.
Hence,
\begin{equation}
\frac{J_{\uparrow \uparrow}-J_{\uparrow \downarrow}}{J_{\uparrow \uparrow}+J_{\uparrow \downarrow}}=
-\rho \sigma_{E} L
\end{equation}

Thus, measurement of the dependence of intensity of passed through the target neutrons on the
orientation of $\overrightarrow{E}$ with respect to the $\overrightarrow{k}$ direction
allows one to get information about T-odd P-even cross section $\sigma_{E}$.

For the above experiment one need not to prepare an aligned target and polarized neutrons,
therefore, there are no restrictions for the type of nuclei. In addition, the target width can be
significantly increased, along with the broadening of the energy range for neutrons.

The considered effect can be caused by either splitting of levels of a nucleus in an electric field
or with levels mixing by this field. Due to this reason, experiments would be better to carry out either by the
use of narrow resonances or near closely set resonances of different parity that let us to 
increase coefficients of levels
mixing by the electric field.

\end{document}